\documentstyle[12pt,a4]{article}

\begin{document}
\title{Evolution of gravitational waves through the 
cosmological QCD transition}
\author{Dominik J. Schwarz \\
        Institut f\"ur Theoretische Physik, Universit\"at Frankfurt, \\
        Postfach 11 19 32, 60054 Frankfurt am Main, Germany
\thanks{e-mail: dschwarz@th.physik.uni-frankfurt.de}}
\date{11 August 1998}
\maketitle

\begin{abstract}
The spectrum of gravitational waves that have been produced in inflation 
is modified during cosmological transitions. Large drops in the number of 
relativistic particles, like during the QCD transition or at 
$e^+e^-$ annihilation, lead to steps in the spectrum of gravitational 
waves. We calculate the transfer function for the differential energy density 
of gravitational waves for a first-order and for a crossover QCD transition. 
\end{abstract}

\section{Introduction}

Detecting a stochastic background of gravitational waves \cite{Allen} would 
open a new window to the early Universe. Primordial gravitational waves are 
predicted to be generated during inflation \cite{Starobinskii} and could 
be detected with upcoming cosmic microwave background observations. Defects, 
like cosmic strings, produce stochastic gravitational waves as well 
\cite{Vilenkin}. Inflation and defects predict an almost scale-invariant 
energy density per logarithmic frequency interval for the most interesting 
frequencies ($\sim 10^{-8}$ Hz for pulsar timing, $\sim 10^{-3}$ Hz for LISA, 
and $\sim 100$ Hz for LIGO) of gravitational waves. 

The aim of this paper is to study the evolution of primordial
gravitational waves through transitions of the equation of state, 
especially the QCD transition. A step in the gravitational wave 
spectrum comes from the large drop of the number of relativistic particles 
(by a factor $\sim 3$) during the QCD transition. Since entropy is  
conserved, the growth rate of the Hubble radius $H^{-1}$ is diminished 
during the transition.
Thus, the rate at which modes cross into the horizon is changed during the
transition and a step in the spectrum shows up at frequencies of order of the 
Hubble rate at the transition.
For the QCD transition this frequency is $10^{-7}$ Hz today and the 
step is a $30\%$ correction. Other large drops in entropy density 
happen at $e^+ e^-$ annihilation (which gives rise to 
a $20 \%$ correction in the energy spectrum) and at a 
GUT phase transition. The typical frequencies are $10^{-10}$ Hz for 
$e^+ e^-$ annihilation and $10^9$ Hz for the GUT transition. 

Similar steps in the differential spectrum have been studied for 
gravitational waves generated by cosmic strings \cite{Vilenkin}.
These gravitational waves are generated on subhorizon scales
when the cosmic strings decay. Their frequency is $> \alpha^{-1} H$, 
where $\alpha$ is the ratio between the typical size of a string loop 
and the Hubble radius at formation of the loop. $\alpha$ is smaller than 
$0.1$ and might be as small as $10^{-5}$ \cite{Caldwell}.  
In this situation steps in the differential spectrum follow from the 
conservation of entropy for decoupled species (gravitons) during cosmological 
phase transitions \cite{Bennett}. 
This interpretation applies to modes that have been inside the horizon 
long before the transition or that have been generated on subhorizon scales.
However, for superhorizon modes
entropy and energy of a gravitational wave are not defined.
We therefore cannot rely on the conservation of entropy argument when 
dealing with gravitational waves from inflation. 

In Sec.~2 we briefly recapitulate the generation of primordial 
gravitational waves from inflation. The equation of state during
the QCD transition is discussed in Sec.~3. In Sec.~4 we explain 
the origin of the step 
in the spectrum and present results from numerical calculations, for a first 
order QCD transition and for a QCD crossover. A further discussion  
of the steps in the spectrum of primordial gravitational waves 
from the  QCD transition and other cosmological transitions, 
especially $e^+ e^-$ annihilation, is presented in Sec.~5.

\section{Primordial spectrum of gravitational waves}

Let us recall the production of gravitational waves during 
inflation \cite{Starobinskii,gwspectrum}. The line element of gravitational 
waves is given by 
${\rm d}s^2 = -{\rm d}t^2 + a^2(\delta_{ij}+h_{ij}){\rm d}x^i{\rm d}x^j$.
$h_{ij}$ is a transverse, traceless tensor. $a$ denotes the scale factor. 
The spatial average $\langle h_{ij}(x)h^{ij}(x+r) \rangle = 
\int j_0(kr) k^3 |h_k|^2 {\rm d}\ln k$ defines the power spectrum
$|h_k|^2$. $k$ is the comoving wavenumber. We define the rms amplitude 
$h$ of a gravitational wave per logarithmic frequency interval: 
$h \equiv k^{3/2}|h_k|$. The linearized equation of motion for $h(t)$ reads
\begin{equation}
\label{ev}
\ddot{h} + 3 H \dot{h} + {k^2\over a^2} h = 0 \ ,
\end{equation}
where the differentiation is taken with respect to cosmic time $t$ and $H
\equiv \dot{a}/a$.

During the quasi-de Sitter period gravitational waves are produced with 
almost scale-invariant spectrum. For superhorizon scales, 
$k_{\rm ph} \equiv k/a \ll H$, the slow roll approximation gives 
\cite{gwspectrum}:
\begin{equation}
\label{h2}
h^2 = {16 \over \pi} F(\epsilon)
\left(H_{\rm dS}\over M_{\rm P}\right)^2_{k=Ha} \ ,
\end{equation}
where $\epsilon \equiv -\dot{H}/H^2 \ll 1$ is the slow roll parameter and
$F(\epsilon) = 1 -(\gamma_{\rm E} + \ln 2 - 1)\epsilon + {\cal O}(\epsilon^2)$;
$\gamma_{\rm E} + \ln 2 - 1 \approx 0.27$. 
The amplitude of gravitational waves stays constant until the second horizon 
crossing and decays as $1/a$ thereafter, 
$h \simeq C_k \sin(k\eta + \delta_k)/a $,
where $\eta = \int {\rm d}t/a$ is conformal time.
$C_k$ and $\delta_k$ are determined by matching to the superhorizon solution.
 
For subhorizon modes, $k_{\rm ph} \gg H$,
the energy density of gravitational waves can be 
defined. The space-time average of the energy-momentum tensor over 
several wavelengths gives $\rho_{\rm g} = - (M_{\rm P}^2/32 \pi) 
\langle \dot{h}_{ij} \dot{h}^{ij}\rangle$. The energy density per logarithmic 
interval in $k$ is related to the rms amplitude $h$: 
\begin{equation}
\label{rhodef}
k{{\rm d}\rho_{\rm g}\over {\rm d}k} = {M_{\rm P}^2 \over 32 \pi}
k_{\rm ph}^2 \frac12 h^2 \ .
\end{equation}
The factor $1/2$ comes from the time average over several oscillations. 
The energy fraction in gravitational waves, per logarithmic interval in $k$, 
is defined by 
\begin{equation}
\Omega_{\rm g} (k) \equiv  k {{\rm d} \rho_{\rm g}\over {\rm d} k}
{1\over \rho_{\rm c}} \ , 
\end{equation}
where $\rho_c \equiv 3 M_{\rm P}^2 H_0^2 /(8\pi)$. Finally, the fractional
energy density per logarithmic frequency interval reads: 
\begin{equation}
\label{rho}
\Omega_{\rm g} (k) \simeq
{2\over 3 \pi} F(\epsilon) \left( H_{\rm dS}\over M_{\rm P}\right)^2 
\left( k_{\rm ph}\over H_0 \right)^2 
\left({a_{\rm hc}\over a_0}\right)^2 \ .
\end{equation}
By $\simeq$ we indicate that we replaced the exact time dependence of $h$
by $a_{\rm hc}/a_0$.
Modes that cross into the horizon \footnote{We define the moment of horizon 
crossing by the condition $k_{\rm ph} = H$.} 
in the matter dominated epoch have $a_{\rm hc}/a_0 = (H_0/k_{\rm ph})^2$ 
and modes that cross into the horizon in the radiation dominated regime
have $a_{\rm hc}/a_0 = H_0/[(1+z_{\rm eq})^{1/2} k_{\rm ph}]$.
From Eq.~(\ref{rho}) we obtain
\begin{equation}
\label{om}
\Omega_{\rm g} (k) \simeq 
{2 \over 3\pi} F(\epsilon)\left(H_{\rm dS}\over M_{\rm P}\right)^2
\left\{
\begin{array}{ll}
(H_0/ k_{\rm ph})^2 &  H_0 < k_{\rm ph} < H_0 (1+z_{\rm eq})^{1/2}\\
(1+z_{\rm eq})^{-1} &  H_0 (1+z_{\rm eq})^{1/2} < k_{\rm ph} < H_{dS}
\end{array}\right. \ .
\end{equation}

For comparison with experimental limits we use the frequency today, 
$f \equiv 2\pi k_{\rm ph}(t_0)$. In inflationary cosmology the strongest 
limit on $\Omega_{\rm g}$ comes from the anisotropies of the cosmic 
microwave background \cite{CMBex}. This limit yields $\Omega_{\rm g}h^2_{50} 
< 3 \times 10^{-10} (H_0/f)^2$ in the frequency range $H_0 < f < 30 H_0$ 
\cite{CMBth,tilt}. From Eq.~(\ref{om}) we have $\Omega_{\rm g}h^2_{50} < 
7 \times 10^{-12}/z_{\rm eq} \approx 1 \times 10^{-15}$ for modes that 
crossed into the horizon during the radiation dominated 
epoch. Direct limits on $\Omega_{\rm g} h^2_{50}$ in the radiation era have 
been obtained by pulsar timing ($ < 4 \times 10^{-7}$ for $f \approx 
4 \times 10^{-9}$ to $4 \times 10^{-8}$ Hz) \cite{pulsarth,pulsar}
and from big bang nucleosynthesis ($ \int \Omega_{\rm g} {\rm d}(\ln f) 
< 10^{-5}$ for $f > H_{\rm BBN} \sim 10^{-9}$ Hz) \cite{Carr}.
Although, inflation predicts an extremely small amount of energy in
gravitational waves at small scales, this is different for other models of
structure formation. Defects might generate a scale-invariant spectrum of
gravitational waves with amplitudes that may be seen in pulsar timing residuals
in the near future. String cosmology \cite{stringcos}
predicts a different energy spectrum,
$\propto f^3$, thus the BBN constraint, which does not constrain the 
inflationary scenarios at all, turns into a severe restriction.

\section{The cosmological QCD transition}

Lattice QCD results indicate that QCD makes a transition from a phase of
free quarks and gluons to the phase of hadrons at a temperature
$T_\star \approx 150$ MeV \cite{lattice}. This implies a Hubble 
radius $R_{\rm H} \sim 10$ km at the transition. Modes that crossed the 
horizon during the QCD transition have frequencies $f_\star \sim 10^{-7}$ 
Hz today. The order of the QCD transition is still a subject of debate.
Lattice QCD with Wilson quarks indicates that the QCD transition
is of first order for the physical values of the quark masses \cite{Iwasaki},
whereas results with staggered quarks \cite{Brown} indicate a crossover for
the physical quark masses. 

We consider both scenarios for the cosmological QCD transition, a 
first-order phase transition and a smooth crossover. A first-order QCD 
transition starts with a short period of supercooling and the nucleation 
of hadron bubbles. The supercooling is tiny, $\delta T/T \sim 10^{-3}$, 
which implies that the entropy production is negligible, 
$\Delta S/S \sim 10^{-6}$ \cite{Witten,SSW}. After enough hadron bubbles have 
been nucleated to reheat the Universe to the critical temperature, further 
bubble formation is suppressed and the bubbles grow adiabatically during
the remaining $99 \%$ of the transition. For our purpose the cosmic fluid 
remains in thermal equilibrium throughout the phase transition. From recent
lattice QCD results \cite{Iwasaki2} for the surface tension and for latent 
heat we find that the typical bubble separation is $\sim 10^{-6} R_{\rm H}$
\cite{SSW}. 

A simple model for a first-order transition is given by the bag model 
\cite{DeGrand}. In this model the entropy density jumps at the critical 
temperature $T_\star$
\begin{equation}
\label{bag}
s(T) = {2 \pi^2\over 45} g_{\rm a} T^3
\left[1 + {\Delta g\over g_{\rm a}}\theta(T - T_\star)\right] \ ,
\end{equation}
where the effective number of relativistic helicity degrees of freedom 
before the transition are $g_{\rm b} = 51.25$ (2 quark flavors, photons, and
leptons\footnote{$e^\pm$, $\mu^\pm$, and all $\nu$s.}) and 
$g_{\rm a} = 17.25$ (pions, photons, and leptons) after
the transition; $\Delta g \equiv g_{\rm b} - g_{\rm a}$. 
At $T_\star$ the high and low temperature phases can coexist. 

For a smooth crossover we assume for the entropy density \cite{Blaizot}
\begin{equation}
\label{crossover}
s(T) = {2 \pi^2\over 45} g_{\rm a} T^3 \left[
1 + \frac12 {\Delta g\over g_{\rm a}}
\left[1 + \tanh\left(T-T_\star\over \Delta T\right)\right]\right] \ .
\end{equation}
We recover the bag model for $\Delta T \to 0$. Both models coincide 
at temperatures far away from $T_\star$. 

In both models the pressure is determined from $p = \int s {\rm d} T$ and 
the energy density follows from the second law of thermodynamics, 
$\rho = sT-p$. The relation between the temperature and the scale factor 
follows from the conservation of entropy,
\begin{equation}
\label{Ta}
{{\rm d}T\over{\rm d}\ln a} = - {3 s\over {\rm d} s/{\rm d} T} \ ,
\end{equation}
except for $T = T_\star$ in the case of a first-order phase transition.
In the bag model $T \propto 1/a$ for $T\neq T_\star$. During a first-order 
QCD transition, i.e.~$T = T_\star$, the pressure $p(T_\star) \equiv p_\star$ 
is constant. The energy density $\rho(a)$ is obtained from the first law of 
thermodynamics $d\rho = -3(\rho + p_\star)da/a$. In the bag model the 
transition lasts a third of a Hubble time. For a detailed 
discussion of the cosmological QCD transition see Ref.~\cite{SSW}. 

\section{Step in the differential energy spectrum}

\begin{figure}
\begin{center}
\input{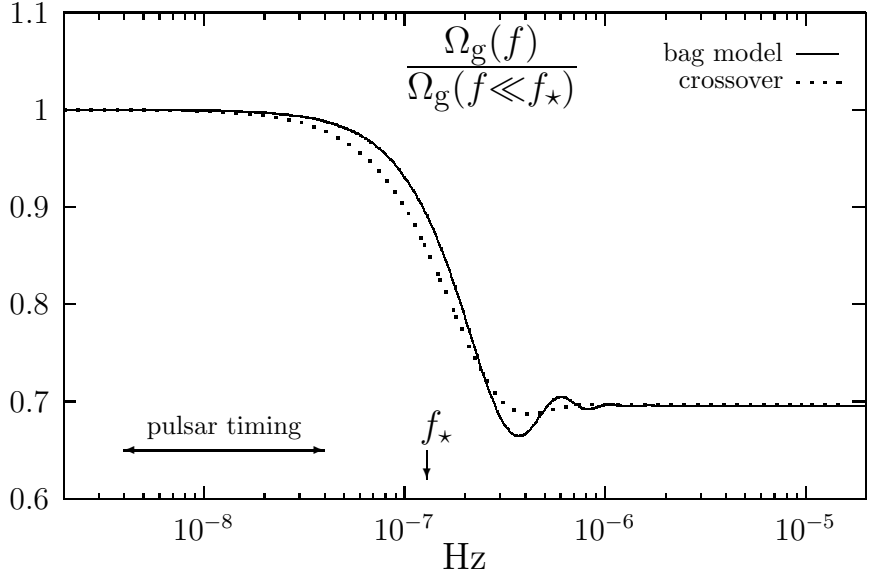}
\end{center}
\caption{The modification of the energy density, per logarithmic frequency
interval, for primordial gravitational waves from the QCD transition.
\label{fig1}}
\end{figure}

Figure \ref{fig1} shows the transfer function $\Omega_{\rm g}(f)/ 
\Omega_{\rm g}(f \ll f_\star)$ from a numerical integration of Eq.~(\ref{ev}). 
The typical frequency scale is 
\begin{equation}
f_\star \approx 1.36 \left(g\over 17.25\right)^{\frac12} 
{T_\star \over 150 \mbox{MeV}} 10^{-7} \mbox{Hz} \ , 
\end{equation}
which corresponds to the mode that crosses the Hubble horizon at the end of 
the bag model QCD transition. Scales that cross into the horizon after the 
transition (l.h.s. of the figure) are unaffected, whereas modes that
cross the horizon before the transition are damped by an additional factor 
$\approx 0.7$. The modification of the differential spectrum has been
calculated for a first-order QCD transition (bag model), Eq.~(\ref{bag}), and 
a crossover QCD transition, Eq.~(\ref{crossover}), with $\Delta T/T_\star=0.3$. 
In both cases the step extends over one decade in frequency. The detailed 
form of the step is almost independent from the order of the transition.
To solve Eq.~(\ref{ev}) numerically, it proofs useful to use the 
scale factor as evolution parameter for the first order transition
and the temperature for the crossover transition. $a$ and $T$ are related by
Eq.~(\ref{Ta}).

The size of the step can be calculated analytically: 
Long before and after the transition, in the radiation dominated era,
\begin{equation}
\label{RH}
H \propto  g^{-1/6} a^{-2} \ , 
\end{equation}
from the Friedmann equation and the conservation of entropy. On the other hand, 
$k_{\rm ph} \propto 1/a$, independently of the equation of state. 
Thus, the rate at which modes enter the Hubble horizon is changing 
during the transition. This change leads to the step in the transfer function.

For a fixed mode $k_{\rm a}$ that crosses the horizon after the phase 
transition the amplitude $h$ is constant during the transition, the transfer
function is one. Now, consider a mode $k_{\rm b}$ that crosses the horizon 
before the transition. Its amplitude decays proportionally to $a_{\rm hc}/a$. 
From Eq.~(\ref{rho}) we find 
\begin{equation}
\label{prop}
\Omega_{\rm g}(k_{\rm b}) \propto k_{\rm ph}^2 a_{\rm hc}^2 \propto 
g_{\rm b}^{-1/3} \ ,
\end{equation}
where we have used Eq.~(\ref{RH}) and $k_{\rm ph} = H$ at horizon crossing. 
The constants of proportionality have been neglected in Eq.~(\ref{prop}),
because they drop out from the transfer function below. Comparing the 
differential energy spectrum for modes that cross into the horizon before 
the transition and for modes that cross into the horizon after the transition 
gives the ratio
\begin{equation}
\label{ratio}
{\Omega_{\rm g}(f \gg f_\star) \over \Omega_{\rm g}(f\ll f_\star)} = 
\left(g_{\rm a}\over g_{\rm b}\right)^{\frac13} \approx 0.696 \ ,
\end{equation}
for the QCD transition, which coincides with the numerical integration in  
Fig.~\ref{fig1}. The size and position of the step in the logarithmic 
spectrum is independent of the order of the transition! 

The result (\ref{ratio}) is in agreement with the entropy conservation
of subhorizon gravitational waves. However, for superhorizon modes
the entropy is not defined.  

Let us estimate the slope of the step in the transfer function.
From the Friedmann equation and the 1st law of thermodynamics we
obtain the Hubble rate as a function of the scale factor ($w \equiv p/
\rho$)
\begin{equation}
{\rm d}\ln H(a) = - \frac32 [1 + w(a)] {\rm d} \ln a \ .
\end{equation}
For $0 < w \leq 1/3$ during the phase transition the Hubble expansion 
changes during the transition, i.e., $H \propto a^{-\beta}$ with $\beta = 
\beta(a)$ in the range $(3/2,2)$. This holds true if the change in 
$w$ is small, i.e.~$({\rm d}w/{\rm d} a) \Delta a \ll 1 + w$. To estimate 
the exponent
\begin{equation}
n(f) \equiv {{\rm d}\ln \Omega_{\rm g}(f)\over {\rm d}\ln f} 
\end{equation}
we use the relation ${\rm d} \ln f = {\rm d} \ln (H a)_{\rm hc} = 
[-(1 + 3 w)/2\ {\rm d} \ln a]_{\rm hc}$ and Eq.~(\ref{rho}). We find 
\begin{equation}
\label{n}
n(f) \simeq - 2 {1 - 3 w(a_{\rm hc}) \over 1 + 3 w(a_{\rm hc})} \ ,
\end{equation}
which takes values between $-2$ and $0$ for $w$ between $0$ and $1/3$. 
From the numerical integration (see Fig.~\ref{fig1}) we find that the minimum
of the exponent is $n \approx - 0.8$ for the bag model, which 
corresponds to an effective value $w_{\rm eff} \approx 0.15$. 
This is consistent with the values $w(a) \in (0.09,1/3)$ that are taken 
during the first-order QCD transition. For the crossover transition
the drop in $w$ is smaller and thus the step is not as steep as in the 
first-order transition.

\section{Discussion}

In Fig.~\ref{fig1} we indicated the frequency range ($\sim 1$ yr$^{-1}$)
in which limits on $\Omega_{\rm g}$ have been reported 
from pulsar timing residuals \cite{pulsar}.
The frequencies where the step of the QCD transition would be visible is of 
the order $0.3$ month$^{-1}$. For pulsar timing the power spectrum of 
gravitational waves is more relevant than the energy spectrum. 
The power spectrum is $\propto \Omega_{\rm g}(f) f^{-5}$. Our results show 
that the power spectrum might deviate from the $f^{-5}$ behavior 
over a whole decade in frequency. Depending on the effective value of 
$w$ during the transition it might be as steep as $f^{-7}$ (for $w = 0$). 
However, a realistic estimate for the QCD transition gives 
$w_{\rm min} \geq 0.1$. Thus, the spectral index of the power spectrum, $n-4$,
lies, according to Eq.~(\ref{n}), between $-6.1$ and $-5$. 

The $30\%$ step in the differential spectrum from the QCD transition 
might be bigger than the tilt of the spectrum of gravitational waves from 
inflation (see \cite{tilt} for a computation of the tilt for various 
inflationary models), e.g.~in chaotic inflation.  

As mentioned in the introduction, similar steps in the spectrum might
occur at $e^+ e^-$ annihilation, where $f_{\rm ann} \sim 10^{-10}$ Hz is the 
typical frequency. Eq.~(\ref{ratio}) is modified for $e^+ e^-$ annihilation,
because neutrinos and photons are decoupled. Taking their difference in 
temperature into account [$T_\nu / T_\gamma = (4/11)^{1/3}$] we obtain
\begin{equation}
{\Omega_{\rm g}(f \gg f_{\rm ann}) \over \Omega_{\rm g}(f\ll f_{\rm ann})} = 
{\left[g_\gamma + {4 \over 11} g_\nu \right]^{\frac43}
\left[g_\gamma + (\frac{4}{11})^{\frac43} g_\nu \right]^{-1}
\over g_{\rm b}^{\frac13}}  \approx 0.8 \ ,
\end{equation}
where $g_\gamma = 2, g_\nu = 5.25,$ and $g_b = 10.75$.
Again, this result is in agreement with the entropy conservation of subhorizon 
gravitational waves.  
 
For the electroweak transition the typical frequency ($\sim 6 \times 
10^{-4}$ Hz) lies in the frequency range of LISA, however the modification of
the spectrum is tiny (the only particle that disappears from 
the radiation fluid is the Higgs particle).

We have studied the evolution of gravitational waves through the QCD 
transition in some detail. Although the detection of primordial
gravitational waves is unlikely in the near future, it is exciting that
the particle content in the early Universe is remembered in the 
spectrum of gravitational waves today. The spectrum contains information about
the relativistic degrees of freedom at a certain epoch of the early Universe
and the transition temperatures of various cosmological transitions. On 
the other
hand, if we assume to know the QCD transition temperature from lattice QCD,
measuring the position of the step in the differential energy spectrum
would tell us whether the gravitational waves have been created during 
inflation or in a more recent epoch of the Universe 
($f_\star \sim 10^{-7}$ Hz from inflation, $>10^{-5}$ Hz from cosmic strings). 

\section*{Acknowledgments}

I thank P. Widerin for helping me with the numerical calculations and
R. Caldwell, V. Mukhanov, C. Schmid, and P. Widerin for comments and 
discussions.
This work has been supported by the Swiss National Science Foundation
and by the Alexander von Humboldt foundation.


\begin{thebibliography}{99}
\bibitem{Allen}Reviews on the generation and detection of stochastic 
         gravitational waves can be found in:
         K. S. Thorne, in {\em 300 Years of Gravitation} 
         (eds. S. Hawking and W. Israel, Cambridge University Press, 
         Cambridge, 1987) pp. 330;
         B. Allen, Les Houches Lecture Notes Sept.~1995, 
         preprint gr-qc/9604033.
\bibitem{Starobinskii}A. A. Starobinskii, Pis'ma Zh. Eksp. Teor. Fiz. 
         {\bf 30}, 719 (1979) [JETP Lett. {\bf 30}, 682 (1979)];
         L. F. Abbott and D. D. Harrari, Nucl. Phys. B {\bf 264}, 487 (1986);
         B. Allen, Phys. Rev. D {\bf 37}, 2078 (1988); 
         V. F. Mukhanov, H. A. Feldman, and R. H. Brandenberger, 
         Phys. Rep. {\bf 215}, 203 (1992).
\bibitem{Vilenkin}A. Vilenkin, Phys. Lett. {\bf 107B}, 47 (1981);
         A. Vilenkin and E. P. S. Shellard, Cosmic Strings and Other 
         Topological Defects (Cambridge University Press, Cambridge, 1994) pp.
         306.
\bibitem{Caldwell}R. R. Caldwell and B. Allen, Phys. Rev. D {\bf 45}, 
         3447 (1992). 
\bibitem{Bennett}D. P. Bennett, Phys. Rev. D {\bf 34}, 3592 (1986); 
         Phys. Rev. D {\bf 34}, 3932 (1986). R. R. Caldwell, R. A. Battye, 
         and E. P. S. Shellard, Phys. Rev. D {\bf 54}, 7146 (1996).
\bibitem{gwspectrum}E. D. Stewart and D. H. Lyth, Phys. Lett. {\bf B} 302,
         171 (1993).
\bibitem{CMBex}C. L. Bennett et al., Astrophys. J. {\bf 464}, L1 (1996).
\bibitem{CMBth}V. A. Rubakov, M. V. Sazhin and A. V. Veryaskin, 
         Phys. Lett. {\bf 115B}, 189 (1982); 
         R. Fabbri and M. D. Pollock, ibid. {\bf 125B}, 445 (1983); 
         L. F. Abbott and M. B. Wise, Nucl. Phys. B {\bf 244}, 541 (1984); 
         A. A. Starobinskii, Pis'ma Astron. Zh. {\bf 11}, 323 (1985)
         [Sov. Astron. Lett. {\bf 11}, 133 (1985)];
         M. S. Turner, M. White, and J. E. Lidsey, Phys. Rev. D {\bf 48},
         4613 (1993).
\bibitem{tilt}M. S. Turner, Phys. Rev. D {\bf 48}, 3502 (1993).
\bibitem{pulsarth}M. V. Sazhin, Astron. Zh. {\bf 55}, 65 (1978) [Sov. Astron.
         {\bf 22}, 36 (1978)]; 
         S. Detweiler, Astrophys. J. {\bf 234}, 1100 (1979).
\bibitem{pulsar}V. M. Kaspi, J. H. Taylor, and M. F. Ryba, 
         Astrophys. J. {\bf 428}, 713 (1994); 
         S. E. Thorsett and R. J. Dewey, Phys. Rev. D {\bf 53}, 3468 (1996);
         M. P. McHugh et al., Phys. Rev. D {\bf 54}, 5993 (1996).
\bibitem{Carr}B. J. Carr, Astron. Astrophys. {\bf 86}, 6 (1980).
\bibitem{stringcos}G. Veneziano, Helv. Phys. Acta {\bf 69}, 553 (1996) and
         references therein.
\bibitem{lattice}For recent reviews see K. Kanaya, Nucl. Phys. B
         (Proc. Suppl.) {\bf 47}, 144 (1996); E. Laermann, Nucl. Phys. B
         (Proc. Suppl.) {\bf 63}, 141 (1998).
\bibitem{Iwasaki}Y. Iwasaki et al., Z. Phys. C {\bf 71}, 343 (1996).
\bibitem{Brown}F. R. Brown et al., Phys. Rev. Lett. {\bf 20}, 2491 (1990).
\bibitem{Iwasaki2}Y. Iwasaki et al., Phys. Rev. D {\bf 46}, 4657 (1992);
        {\bf 49}, 3540 (1994);
        B. Beinlich, F. Karsch, and A. Peikert, Phys. Lett. B {\bf 390}, 268
        (1997).
\bibitem{Witten}E. Witten, Phys. Rev. D {\bf 30}, 272 (1984); 
         J. Ignatius et al., Phys. Rev. D {\bf 49}, 3854 (1994);
         {\bf 50}, 3738 (1994);
         M. B. Christiansen and J. Madsen, Phys. Rev. D {\bf 53},
         5446 (1996).
\bibitem{SSW}C.~Schmid, D.~J.~Schwarz, and P.~Widerin,
         Phys.\ Rev.\ Lett.\ {\bf 78}, 5468 (1997); astro-ph/9807257 (1998).
\bibitem{DeGrand}T. DeGrand and K. Kajantie, Phys. Lett. {\bf 147B},
         273 (1984).
\bibitem{Blaizot}J. P. Blaizot and J. Y. Ollitrault, Phys. Rev. D {\bf 36},
         916 (1987).
\end{thebibliography}
\end{document}